\documentclass{jfm}

\usepackage{graphicx}
\usepackage{newtxtext}
\usepackage{newtxmath}
\usepackage{natbib}
\usepackage{hyperref}
\hypersetup{
    colorlinks = true,
    urlcolor   = blue,
    citecolor  = black,
}

\newcommand{\RomanNumeralCaps}[1]
\linenumbers

\title{Quantitative thermodynamic analyses of nucleation, evolution and stabilization of surface nanobubbles}

\author{Lili Lan\aff{1},
  Yongcai Pan\aff{2},
  Liang Zhao\aff{3,}
  \corresp{\email{zhaoliang@yzu.edu.cn}},
 \and Binghai Wen\aff{1,2,}
 \corresp{\email{oceanwen@gxnu.edu.cn}}}

\affiliation{
\aff{1}College of Physical Science and Technology, Guangxi Normal University, Guilin 541004, China
\aff{2}Key Lab of Education Blockchain and Intelligent Technology, Ministry of Education, Guangxi Normal University, Guilin 541004, China
\aff{3}College of Physical Science and Technology, Yangzhou University, Jiangsu, 225009, China}

\begin{document}
\maketitle

\begin{abstract}
Surface nanobubbles are complex micro- and nanoscale fluid systems. While thermodynamics is believed to dominate nanobubble dynamics, the precise mechanism by which nanobubble evolution is driven by thermodynamics remains unclear. It is essential to understand how nanobubble nucleation and growth, nanoscale contact line movement, and gas diffusion across the liquid-bubble interface are simultaneously driven by the change in free energy, leading to the ultimate thermodynamic equilibrium of surface nanobubble systems. In this paper, we first propose a quantitative theoretical model to elucidate the thermodynamic dominance behind the dynamics and stability of the fluid system with surface nanobubbles. The present model demonstrates that thermodynamic non-equilibrium drives the gas diffusion and the contact line motion of surface nanobubbles. Overcoming the nucleation energy barrier is crucial for bubble nucleation and growth. Surface nanobubbles evolve towards the reduction of the system's free energy and stabilize at the state with minimum free energy. The thermodynamic equilibrium is accompanied by the mechanical equilibrium at the contact line and the gas diffusion equilibrium at the liquid-bubble interface, and the theoretical results are in excellent agreement with the nanobubble morphology observed in experiments. The study highlights the significant influence of gas properties and ambient conditions in promoting bubble nucleation and stability.
\end{abstract}


\section{Introduction}
\label{sec:headings}

Surface nanobubbles (SNBs) are spherical cap-shaped nanoscopic gaseous domains that form at the solid-liquid interface, representing a complex micro- and nanoscale fluid system \citep{Lohse2015,Tan2021,Zhou2021}. Since being experimentally confirmed, the remarkable stability of SNBs has drawn significant interest from interfacial scientists and engineers. This exceptional stability contradicts the widely accepted Epstein-Plesset theory \citep{Epstein1950}, which predicts that sub-micron nanobubbles would dissolve within microseconds. Understanding why SNBs can survive for days and weeks, and how they grow or shrink, is crucial for studying and utilizing nanobubbles in both experiments and industrial applications \citep{Zhang2014,Lohse2018}. Essentially, the stability of SNBs is a thermodynamic issue. Although some pioneering studies have explained the nanobubble stability from a dynamic viewpoint \citep{Lohse2015Pinning,Tan2018,Wen2022}, a comprehensive understanding of the nucleation, evolution and stabilization of surface nanobubbles requires a thermodynamic perspective.

A comprehensive review \citep{Tan2021} highlights the challenge of resolving the stability of nanobubbles within a thermodynamic framework. These problems have been attempted to be solved by examining the conditions for the spontaneous growth of bubbles based on the second law of thermodynamics. \citet{Liu2013,Liu2018} proved the thermodynamic stability of pinned nanobubbles using lattice density functional theory in the grand canonical ensemble. Attard's theoretical analysis \citep{Attard2015,Attard2016} showed that only the pinned surface bubbles can minimize the system's free energy at the critical radius. \citet{Gadea2023} ingeniously used a finite-sized hydrophobic disk as the substrate. Their results demonstrate that when the contact line slides to the disk's edge and becomes pinned, the pinned bubble minimizes the system's free energy in an open environment. Regarding the stability of unpinned nanobubbles, \citet{Ward1984} in their earlier study showed that the stable state of bubble nuclei is a result of mutual competition with the dissolved gas in the liquid phase within a closed volume. Then, \citet{Zargarzadeh2016,Zargarzadeh2019} addressed the free energy of SNBs as a function of size, identifying thermodynamically stable and metastable states for bubbles ranging from hundreds of nanometers to a few micrometers. These treatments examine the nucleation of multiple identical nanobubbles in a container with a finite amount of gas molecules, and also considers a realistic case where both the solute gas and the solvent are present in the liquid and bubble phases. In the above studies, gas diffusion and nanobubble dynamics processes have yet to be considered. 

Scientists explain the dynamic stability of surface nanobubbles (SNBs) by primarily focusing on the behavior of the three-phase contact line and the gas flow at the liquid-bubble interface. The main characteristics of the movement of bubble contact lines include: pinning \citep{Liu2013}, self-pinning \citep{Wang2009,Ren2020,Guo2019,Chen2020} and unpinning \citep{Petsev2020}. Recently, \citet{Petsev2020} considered that the adsorption of gas molecules to the substrate lowered the solid-gas interface energy and explained the flat morphology of nanobubbles from a thermodynamic perspective. The bubble-liquid interface was found to be gas permeable by \citet{German2014}, the gas diffusion through the interface drives the growth or shrinkage of SNBs. \citet{Brenner2008} suggested a dynamic equilibrium mechanism that the gas attracted by the hydrophobic walls influx at the contact line can compensate for the gas outflow. The gas diffusion from a surface nanobubble is analogous to the evaporation of a pinned drop, which was exactly solved by \citet{Popov2005}. Incorporating Henry’s law into Popov’s solution, \citet{Lohse2015Pinning} derived the mass change of a pinned nanobubble in a supersaturated liquid. Then, \citet{Tan2018} introduced the potential into the pinning model and further explained the experimental observations that surface nanobubbles could survive in undersaturated environments \citep{An2016,Qian2018}. Moreover, experimental phenomena manifest that contact line pinning is not strictly necessary for the mechanical equilibrium of SNBs \citep{Nag2021,Bull2018}. \citet{Wen2022} proposed dynamic models for SNBs on homogeneous and heterogeneous surfaces, which can describe the state transition from the stable nanobubbles to unstable microbubbles \citep{Pan2022,Lan2025}.

In this study, we integrate a non-equilibrium drive for gas diffusion with an analysis of the system’s free energy changes, present quantitative thermodynamic calculations to investigate the critical nucleation, evolution direction, and stable state of SNBs in a closed environment. The gas adsorption effect at the solid-gas interface is introduced. Furthermore, the SNB morphology is examined for several common types of gases, taking into account the different actual gas properties.  

\section{Model}
\label{sec:headings}

\begin{figure}
    \centering
    \includegraphics[width=12cm]{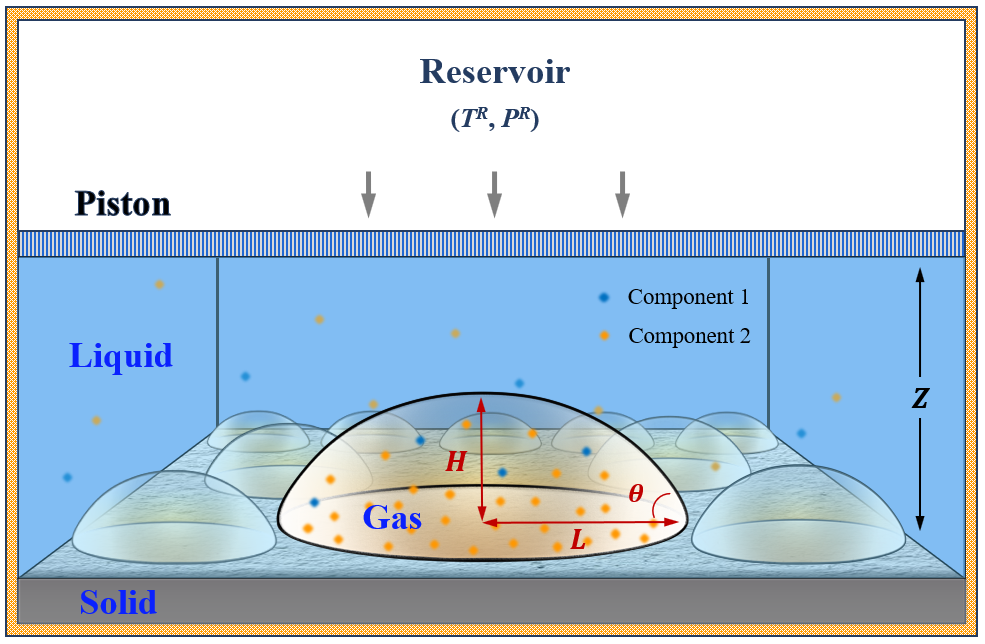}
    \captionsetup{justification=justified, format=plain, width=\textwidth}
    \caption{Schematic of closed fluid system with nanobubbles formed on a smooth solid surface submerged in a liquid-gas solution. The nanobubble morphology is characterized by three geometric parameters: footprint radius $L$, height $H$ and bubble-side contact angle $\theta$. The reservoir has the constant temperature $T^{\mathrm{R}}$ and the constant pressure $P^{\mathrm{R}}$. A movable piston seals the liquid and keeps its pressure at the same level as the reservoir. The area of the solid substrate is $A^{\mathrm{S}}$, and Z represents the initial height of the liquid phase. The liquid contains two types of particles: component 1 and component 2.}
    \label{Fig.1}
\end{figure} 
 
An isothermal-isobaric ensemble (NPT) is provided for the quantitative thermodynamic analysis of the SNB system. As shown in \autoref{Fig.1}, the SNB system consists of the solid, liquid, and gas (bubble) phases, as well as the interfaces between them \citep{Ward1984,Zargarzadeh2016}. The SNB system is in contact with a reservoir by means of a movable piston, which seals it off from the reservoir, and maintains its temperature and pressure at the same level as to the reservoir, thus creating a closed, constant temperature and pressure environment. 

The closed SNB system, together with the external reservoir, can be considered as an isolated system in which the total internal energy, total volume and total particle number are conserved. According to the second law of thermodynamics, the total entropy of the system increases for any change towards equilibrium. Thus, the system's free energy $B$, also known as the thermodynamic potential, can be expressed by \citep{Ward1984,Zargarzadeh2016}
\begin{equation}
B=G^{\mathrm{L}}+F^{\mathrm{G}}+F^{\mathrm{LG}}+F^{\mathrm{SL}}+F^{\mathrm{SG}}+P^{\mathrm{L}} V^{\mathrm{G}},
\label{eq.1}
\end{equation}
where G and F denote the Gibbs and Helmholtz functions, $P$ and $V$ denote the pressure and the sum of SNB volume. The superscripts R, L, G, LG, SL, and SG, represent the reservoir, liquid, gas (bubble), liquid-gas, solid-liquid, and solid-gas phases, respectively. Equation (\ref{eq.1}) indicate that the free energy for the SNB system is not solely one of the conventional thermodynamic potentials, the detailed process is presented in the supplementary materials.

This study considers two components in the closed liquid environment: water ($\text{H}_2\text{O}$) as component 1, and a specific gas that can dissolve in the water as component 2, denoted by subscripts. Therefore, Equation (\ref{eq.1}) can be calculated as
\begin{equation}
\begin{aligned}
B= & \left(\mu_1^{\mathrm{L}} n_1^{\mathrm{L}}+\mu_2^{\mathrm{L}} n_2^{\mathrm{L}}\right)+\left(\mu_1^{\mathrm{G}} n_1^{\mathrm{G}}+\mu_2^{\mathrm{G}} n_2^{\mathrm{G}}-P^{\mathrm{G}} V^{\mathrm{G}}\right)+ \left(\mu_1^{\mathrm{LG}} n_1^{\mathrm{LG}}+\mu_2^{\mathrm{LG}} n_2^{\mathrm{LG}}+\gamma^{\mathrm{LG}} A^{\mathrm{LG}}\right)+\\
& \left(\mu_1^{\mathrm{SL}} n_1^{\mathrm{SL}}+\mu_2^{\mathrm{SL}} n_2^{\mathrm{SL}}+\gamma^{\mathrm{SL}} A^{\mathrm{SL}}\right)+ \left(\mu_1^{\mathrm{SG}} n_1^{\mathrm{SG}}+\mu_2^{\mathrm{SG}} n_2^{\mathrm{SG}}+\gamma^{\mathrm{SG}} A^{\mathrm{SG}}\right)+P^{\mathrm{L}} V^{\mathrm{G}} ,
\end{aligned}
\label{eq.2}
\end{equation}
where $\mu$ is chemical potential, $n$ is the molecular number, $\gamma$ is the interface tension, and $A$ is the interface area. The SNB system in a completely liquid state, i.e. all the gas is dissolved in the liquid, is chosen as the reference, with relevant parameters denoted by subscript "0". So that the free energy of the system with respect to the reference can be defined by
\begin{equation}
\begin{aligned}
\Delta B= & \gamma^{\mathrm{LG}}\left(A^{\mathrm{LG}}-A^{\mathrm{SG}} \cos \theta\right)-\left(P^{\mathrm{G}}-P^{\mathrm{L}}\right) V^{\mathrm{G}}+N_1\left(\mu_1^{\mathrm{L}}-\mu_{10}\right)+ \\
& N_2\left(\mu_2^{\mathrm{L}}-\mu_{20}\right)+\left(n_2^{\mathrm{G}}+n_2^{\mathrm{LG}}+n_2^{\mathrm{SG}}\right)\left(\mu_2^{\mathrm{G}}-\mu_2^{\mathrm{L}}\right),
\end{aligned}
\label{eq.3}
\end{equation}
where $N_1$ and $N_2$ represent the total number of water and specific gas molecules in the SNB system. Compared to Ward's thermodynamic model, our study comprehensively explores the process of the system state transition from thermodynamic non-equilibrium to equilibrium. As depicted in Equation (\ref{eq.3}), the system's dynamic evolution is propelled by the chemical potential difference $\mu_2^{\mathrm{G}}-\mu_2^{\mathrm{L}}$ between the component 2 gas in bubbles and liquid, resulting in gas diffusion at the bubble-liquid interface. This process drives the growth or shrinkage of surface nanobubbles. It is clear that when the system reaches equilibrium, $\mu_2^{\mathrm{G}}=\mu_2^{\mathrm{L}}$ and the surface nanobubbles become stable. In this spontaneous process, the system's free energy decreases, allowing the evolution to continue until $\Delta B$ reaches a minimum that satisfies the given constraints. The criterion for the system being in a thermodynamic equilibrium state is that $\Delta B$ is a minimum \citep{Ward1984,Zargarzadeh2016}.

Introducing the calculation of the chemical potential in each phase, Equation (\ref{eq.3}) can be further expressed as follows:
\begin{equation}
\begin{aligned}
\Delta B= & \gamma^{\mathrm{LG}}\left(A^{\mathrm{LG}}-A^{\mathrm{SG}} \cos \theta\right)-\left(P^{\mathrm{G}}-P^{\mathrm{L}}\right) V^{\mathrm{G}}+N_1 k_{\mathrm{B}} T\left(\frac{N_2}{N_1}-\frac{n_2^{\mathrm{L}}}{n_1^{\mathrm{L}}}\right)+ \\
& N_2 k_{\mathrm{B}} T \ln \left(\frac{n_2^{\mathrm{L}}}{n_{2 {\mathrm{s}}}^{\mathrm{L}}} \frac{N_{2 {\mathrm{s}} 0}}{N_{20}}\right)+\left(n_2^{\mathrm{G}}+n_2^{\mathrm{LG}}+n_2^{\mathrm{SG}}\right) k_{\mathrm{B}} T \ln \left(\frac{P_2^{\mathrm{G}}}{P^{\mathrm{L}}} \frac{n_{2\mathrm{s}}^{\mathrm{L}}}{n_2^{\mathrm{L}}}\right),
\end{aligned}
\label{eq.4}
\end{equation}
where $k_{\mathrm{B}}$ is the Boltzmann constant, the subscript "s" indicates a gas-saturated state in liquid. Equation (\ref{eq.4}) shows that the change in the system's free energy is influenced by the real-time morphology of SNBs and the molecular distributions in each phase. The nanobubble shape is determined by the mechanical equilibrium at three-phase contact line and the gas molecular number in the bubble, which is adjusted by the gas diffusion across the liquid-bubble interface. The thermodynamic non-equilibrium drives the gas diffusion and the contact line motion of surface nanobubbles, and it is crucial for the quantitative thermodynamic analyses to calculate the gas diffusion rate and the nanobubble morphology in real time.

The distribution of the component 2 gas within the bubbles follows ideal gas properties \citep{Ward1984,Zargarzadeh2016,Attard2016,Petsev2020}. We consider the Langmuir adsorption effect \citep{Szyszkowski1908,Swenson2019} of the component 2 gas molecular on the solid surface, thus the adsorbed molecular number is calculated by
\begin{equation}
n_2^{\mathrm{SG}}=\frac{A^{\mathrm{SG}}}{b}\left(\frac{K P_2^{\mathrm{G}}}{1+K P_2^{\mathrm{G}}}\right),
\label{eq.5}
\end{equation}
where $b$ is the cross-sectional area of a single adsorbing molecule, and $K$ is the equilibrium adsorption constant in units of inverse pressure, and $P_2^{\mathrm{G}}$ is the partial pressures of component 2 gas in bubble. The gas adsorption generally lowers the solid-gas interface tension, and then the Young-Laplace equation needs to be modified by the Szyszkowski equation \citep{Szyszkowski1908} which is thermodynamically equivalent to the Langmuir adsorption isotherm \citep{Swenson2019}. The gas-side contact angle with the mechanical equilibrium, as well as the height and footprint radius, can be given by the implicit equation,
\begin{equation}
\cos \theta=\cos \mathit{\Theta}+\frac{k_{\mathrm{B}} T}{b \gamma^{\mathrm{LG}}} \ln \left(\frac{1+K P_2^{\mathrm{G}}}{1+K P^{\mathrm{L}}}\right),
\label{eq.6}
\end{equation}
where $\mathit{\Theta}$ is the macroscopic gas-side contact angle that symbolizes the degree of wettability of substrate surface ($\mathit{\Theta}<90^{\circ}$ is hydrophobic, otherwise is hydrophilic). 

Gas diffusion in the liquid can be considered as a quasistatic limit, and the gas concentration field $c$ around the SNBs follows the quasistatic diffusion equation $\partial_t c=D \nabla^2 c \approx 0$, where $D$ is the diffusion constant \citep{Attard2013}. \citet{Lohse2015Pinning} combined Popov's equation \citep{Popov2005} and Henry’s law to calculate the diffusion rate of a pinned surface nanobubble. Molecular dynamics simulations have shown that the timescale for molecules' thermal motion to equilibrate a contact angle is within nanoseconds \citep{Guo2019,Wang20092,Maheshwari2016}, which is more than three orders of magnitude faster than the diffusion timescale. From the perspective of diffusion, surface nanobubbles always retain the mechanical equilibrium during dynamic growth or shrinkage. Therefore, \citet{Wen2022} constrained the dynamic bubble morphology changed by the gas diffusion during the bubble evolution on a homogeneous substrate. By introducing Equation (\ref{eq.6}), the diffusion rate of component 2 gas molecules within the SNBs is derived as
\begin{equation}
\frac{d n_2^{\mathrm{G}}}{d t}=-\frac{\pi D c_{\mathrm{s}}}{m} L\left(\frac{2 \gamma^{\mathrm{LG}}}{H P^{\mathrm{L}}}\left(1-\cos \Theta-\frac{k_{\mathrm{B}} T}{b \gamma^{\mathrm{LG}}} \ln \left(\frac{1+K P_2^{\mathrm{G}}}{1+K P^{\mathrm{L}}}\right)\right)-\zeta(H, \theta)\right) f(\theta),
\label{eq.7}
\end{equation}
where $\zeta(H, \theta)$ represents the dynamic gas oversaturation in the liquid during bubble evolution. $c_{\mathrm{s}}$ is the solubility of gas, $m$ is the gas molar mass, and $f(\theta)$ is a geometric term \citep{Popov2005}. When the gas diffusion reaches equilibrium, i.e. $d n_2^{\mathrm{G}} / d t=0$, the SNBs reach a stable state.

Equation (\ref{eq.6}) imposes mechanical equilibrium constraints on the three-phase contact line, controlling the shape changes of the SNBs; while Equation (\ref{eq.7}) describes the diffusion direction and rate of gas molecules at the gas-liquid interface, driving the growth and contraction of the SNBs. Therefore, by substituting the changes in the particle number in each phase and the accompanying changes in bubble morphology into Equation (\ref{eq.4}), we can quantitatively analyze the dynamic system's free energy changes during the nucleation, growth, contraction, and stabilization of SNBs.

\section{Results and discussion}
\label{sec:headings}

We consider a nanobubble system, in which $q=1 \times 10^{12}$ nanobubbles are evenly distributed on a surface with $A^{\mathrm{S}}=1 \mathrm{~m}^{2}$. The initial height of the liquid phase is $Z=1 \mathrm{~\mu m}$. The reservoir keeps at a constant room temperature $T^{\mathrm{R}}=25^{\circ} \mathrm{C}$ and the standard atmospheric pressure $P^{\mathrm{R}}=1\mathrm{~atm}$.  

\subsection {Dynamics and stability of surface nanobubbles}

\begin{figure}
    \centering
    \includegraphics[width=12cm]{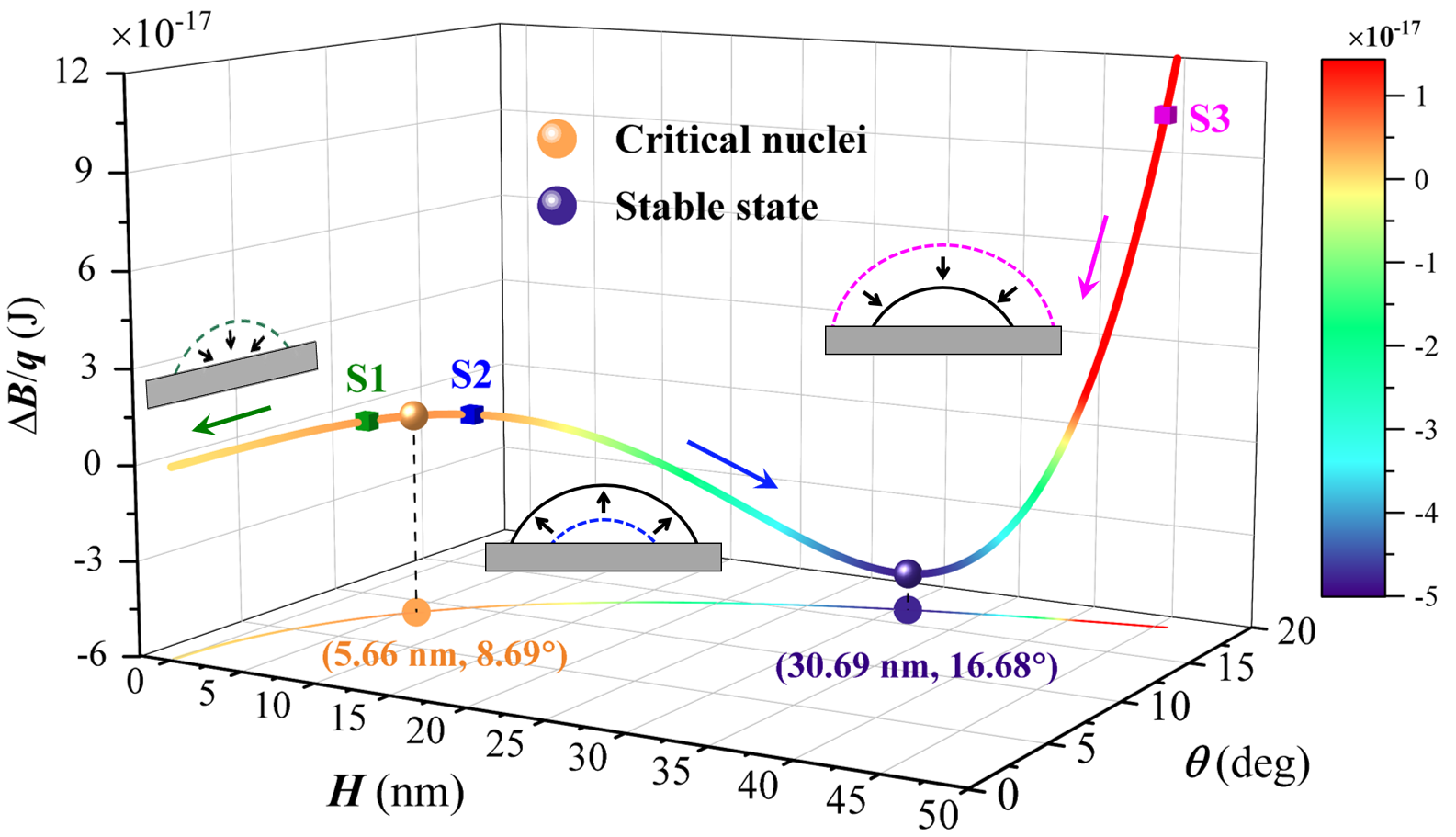}
    \captionsetup{justification=justified, format=plain, width=\textwidth}
    \caption{3D image of system's free energy versus surface nanobubble size (height $H$ and contact angle $\theta$). The symbols of maximum (orange sphere) and minimum (dark purple sphere) free energy corresponds to critical nucleation and equilibrium state of a surface nanobubble, respectively. Three sample points, S1(green square), S2(blue square) and S3(magenta square), are taken as the initial bubble states.}
    \label{Fig.2}
\end{figure} 

In this section, Nitrogen ($\text{N}_2$) is considered as the component 2 gas for quantitative analysis, with $\gamma^{\mathrm{LG}} =0.072\mathrm{~N\cdot m^{-1}}$, $D=2\times 10^{-9} \mathrm{~m^{2}/s}$, and  $b=7.548\times 10^{-2} \mathrm{~nm^2}$ \citep{Lohse2015Pinning,Petsev2020}. By solving Equation (\ref{eq.4}), the relationship between the system free energy and nanobubble sizes can be obtained. The 3D curve of system's free energy change presented in \autoref{Fig.2} illustrates the average impact of each SNB size (height $H$ and contact angle $\theta$) on the system free energy $\Delta B/q$, with gas adsorption strength $K=5\times 10^{-7}$,  gas oversaturation $\zeta=3$ and surface macroscopic gas-side contact angle $\mathit{\Theta}=27^{\circ}$. In the $H-\theta$ axis plane, as the bubble grows, the height $H$ and contact angle $\theta$ show a positive correlation. As shown in \autoref{Fig.2}, as the bubble height $H$ and size increases, gas molecules in the liquid continuously flow into the bubble phase, resulting in a trend of initially increasing and then decreasing and then increasing system free energy, the change in the system's free energy exhibits a maximum state (represented by the orange sphere) and a minimum state (represented by the dark purple sphere). In thermodynamic description, the maximum of the system's free energy is regarded as the critical nucleation energy barrier for surface bubbles, corresponding to the bubble's critical nucleation size projected onto the $H-\theta$ axis plane. Only when the bubble grows to that size and overcomes the corresponding critical energy barrier, can the bubble continue to grow, otherwise it will dissolve. The system with the minimum of free energy is considered to be the most stable state. For the SNB system, the point of minimum free energy in \autoref{Fig.2} should correspond to the most stable size of the surface nanobubble. Then, we take three sample points, S1(green square), S2(blue square), S3(magenta square), as the initial states of SNBs to analyze their evolution.

\begin{figure}
    \centering
    \includegraphics[width=12cm]{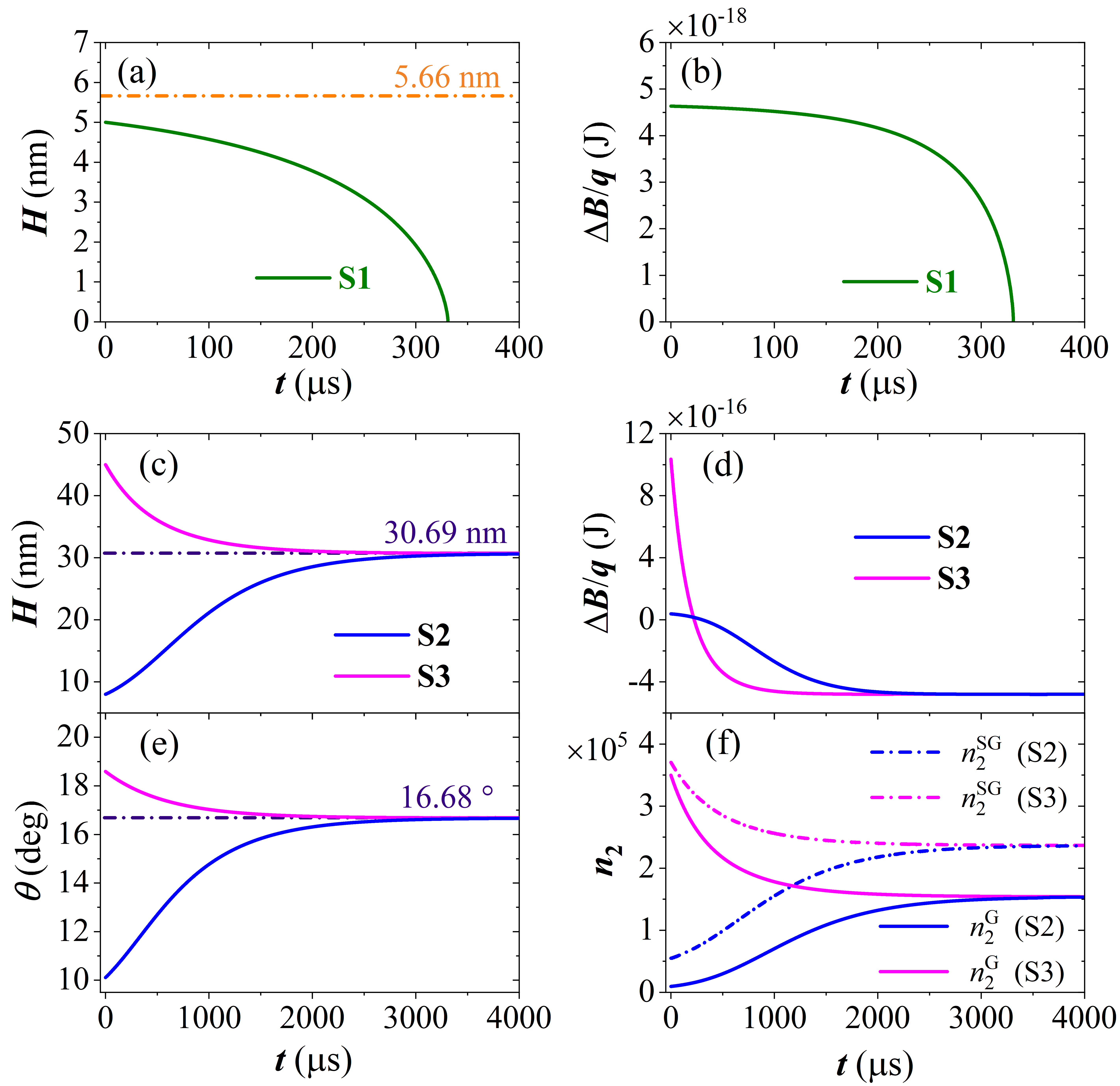}
    \captionsetup{justification=justified, format=plain, width=\textwidth}
    \caption{Evolution of (a) height $H$ of S1 nanobubble, and (b) the corresponding system's free energy $\Delta B/q$ over time. Evolution of (c) height $H$ of S2 and S3 nanobubbles, (d) the corresponding system's free energy $\Delta B/q$, (e) contact angle $
    \theta$, and (f) molecular number $n_2$ of component 2 gas in bubble phase $n_2^{\mathrm{G}}$ and at adsorbed phase $n_2^{\mathrm{SG}}$ over time.}
    \label{Fig.3}
\end{figure} 

In \autoref{Fig.2}, the orange sphere marking the free energy maximum point projected onto the $H-\theta$ axis plane corresponds to the SNB size of $H=5.66\mathrm{~nm}$ and $\theta=8.69^{\circ}$. The initial bubble size in S1 is $H=5\mathrm{~nm}$ and $\theta=7.42^{\circ}$, it is smaller than the critical nucleation size. The dynamic processes of SNBs can be accurately calculated by using gas diffusion Equation (\ref{eq.7}). \autoref{Fig.3}(a) illustrates that S1 nanobubbles which have not yet reached the critical nucleation height are incapable of nucleating and growing, instead shrinking over time until they fully dissolve. In this process, the system's free energy continuously decreases, eventually returning to the reference state with no bubbles forming, as shown in \autoref{Fig.3}(b).

In \autoref{Fig.2}, the dark purple sphere marks the minimum value of the system free energy $\Delta B/q=-4.80\times 10^{-16}\mathrm{~J}$, and the corresponding SNB size projected onto the $H-\theta$ axis plane is $H=30.69\mathrm{~nm}$, $\theta=16.68^{\circ}$. The initial sizes of nanobubbles in S2 and S3 are taken on either side of the bubble size corresponding to the minimum value of free energy cure in \autoref{Fig.2}. The blue curves in \autoref{Fig.3}(c) $\sim $ (f) depict a growing S2 nanobubble starting at 8 nm height, while the magenta curves show the shrinking S3 nanobubble starting at 45 nm height. The changes in the blue curves indicate that as gas molecules continuously flow into the bubble from the liquid, the height $H$ and contact angle $\theta$ of the S2 SNBs increase with time. Conversely, the magenta curves show that as gas molecules flow out of the bubble, the height and contact angle of the S3 SNBs decrease with time. \autoref{Fig.3}(d) illustrates that nanobubbles evolve in a direction that reduces the system's free energy, and they eventually reach a stable state where the evolution curve becomes horizontal, indicating equilibrium in gas diffusion. This stable state aligns with the state of system's free energy minimum value depicted in \autoref{Fig.2}. This evolution phenomenon indicates that the stable state of SNBs is characterized by minimizing the system's free energy. Any deviation in bubble shape will eventually lead it back to the stable state as gas diffuses. In addition, \autoref{Fig.3}(f) shows that when gas diffusion reaches equilibrium, the number of gas molecules adsorbed at the solid-bubble interface (dashed line) exceeds that in the bubble phase (solid line), indicating an ultra-dense aggregation of gas molecules at the bubble-substrate interface. This phenomenon is commonly observed in many molecular dynamic simulations \citep{Chen2018,Yen2021}.

Our thermodynamic model is capable of calculating the changes in system's free energy at any moment and state during the SNB evolution, caused by the gas diffusion. The analyses indicate that the evolution of SNBs consistently progresses towards reducing the system's free energy. When the bubble evolution reaches a stable equilibrium and the gas molecule flux at the bubble-liquid interface becomes zero, the system's free energy is minimized. The result highlights that a stable SNB must satisfy the unification of mechanical equilibrium at three-phase contact line, gas diffusion equilibrium at bubble-liquid interface, and system's thermodynamic equilibrium.

\subsection{Nucleation and stability of nanobubbles in response to environmental conditions}

\begin{figure}
    \centering
    \includegraphics[width=12cm]{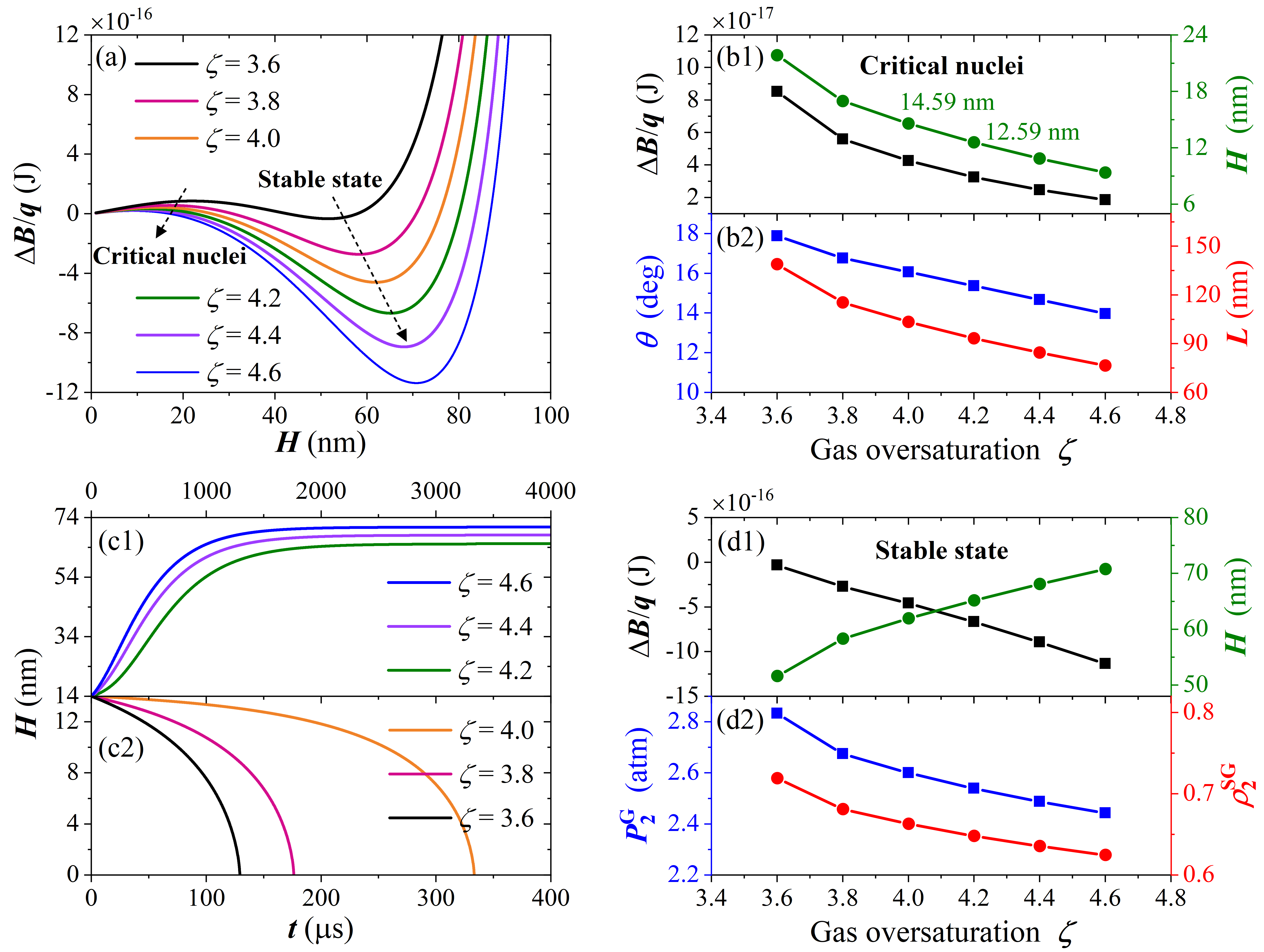}
    \captionsetup{justification=justified, format=plain, width=\textwidth}
    \caption{(a) System's free energy versus bubble height $H$ for different gas oversaturation $\zeta$. (b1) Energy barrier and bubble heigh $H$, (b2) contact angle $\theta$ and footprint radius $L$ of surface nanobubble in the critical nucleation state versus gas oversaturation. (c1) (c2) Evolution of surface nanobubbles starting at 14 nm height over time for different gas oversaturation. (d1) System's free energy and bubble heigh $H$, (d2) gas partial pressure $P_2^{\mathrm{G}}$ in bubble phase and adsorbed gas molecular density $\rho_2^{\mathrm{SG}}$ at solid-bubble interface of surface nanobubble in the stable state versus gas oversaturation.}
    \label{Fig.4}
\end{figure}

In experiments, methods such as electrolysis and alcohol-water exchange \citep{Seddon2011,Zhang2023} are commonly used to adjust the gas concentration in the solution in order to supersaturate or locally supersaturate it. This leads the aggregation of target gas molecules on the solid surface, forming surface nanobubbles. Recent nucleation rate experiments have revealed that the critical nucleus size for heterogeneous nanobubble formation ranges from 5 to 35 nm, much smaller than the bubbles observed by microscopy and studied theoretically \citep{Attard2016,Yatsyshin2021,Yang2013}. \autoref{Fig.4} investigates how different levels of gas oversaturation (with $\mathit{\Theta}=25^{\circ}$ and $K=2\times 10^{-7}$) affect the nucleation and stability of surface nanobubbles. As shown in \autoref{Fig.4}, the system's free energy curve gradually shifts downward with increasing gas oversaturation $\zeta$ in the liquid. \autoref{Fig.4}(b1) and \autoref{Fig.4}(b2) further demonstrate the alterations in the critical nucleation energy barrier and critical nucleation size of SNBs for different $\zeta$ values. The dot-line trends indicate that increasing gas concentration in the solution can lower the critical nucleation energy barrier and decrease the bubble nucleation size, thereby promoting the formation of surface nanobubbles. In the case of sufficient gas oversaturation, the nucleation barrier can disappear, leading to spontaneous nucleation of bubbles. For the stable state of SNBs, \autoref{Fig.4}(d1) illustrates that increased gas concentration leads to a lower minimum value of the system's free energy, resulting in larger stable nanobubbles. Additionally, \autoref{Fig.4}(d2) shows that as the bubble size increases, the gas partial pressure $P_2^{\mathrm{G}}$ in the bubble phase gradually decreases, resulting in a decrease in the adsorbed gas molecular density $\rho_2^{\mathrm{SG}}$ at the solid-bubble interface and a weakening of the gas adsorption strength.

In \autoref{Fig.4}(c1) and \autoref{Fig.4}(c2), the initial height of nanobubbles on the solid surface is 14 nm, which is between the critical nucleation heights of 14.59 nm at $\zeta=4$ and 12.59 nm at $\zeta=4.2$. This indicates that in the liquid conditions with a gas oversaturation of $\zeta=4$ or lower, the initial nanobubble size does not reach the nucleation size. \autoref{Fig.4}(c2) illustrates that bubbles gradually shrink and dissolve in the liquid over time, with gas diffusing outward faster as the gas oversaturation decreases. Conversely, under the liquid condition with gas oversaturation of $\zeta=4.2$ or higher, as shown in \autoref{Fig.4}(c1), nanobubbles continue to grow and eventually stabilize at an equilibrium state. The growth rate of nanobubbles increases with higher gas oversaturation levels. \autoref{Fig.4} demonstrates that increasing gas oversaturation not only lowers the critical nucleation barrier of bubbles, but also reduces the system's free energy in the stable state, allowing SNBs to survive more stably without being easily destroyed.

\begin{figure}
    \centering
    \includegraphics[width=12cm]{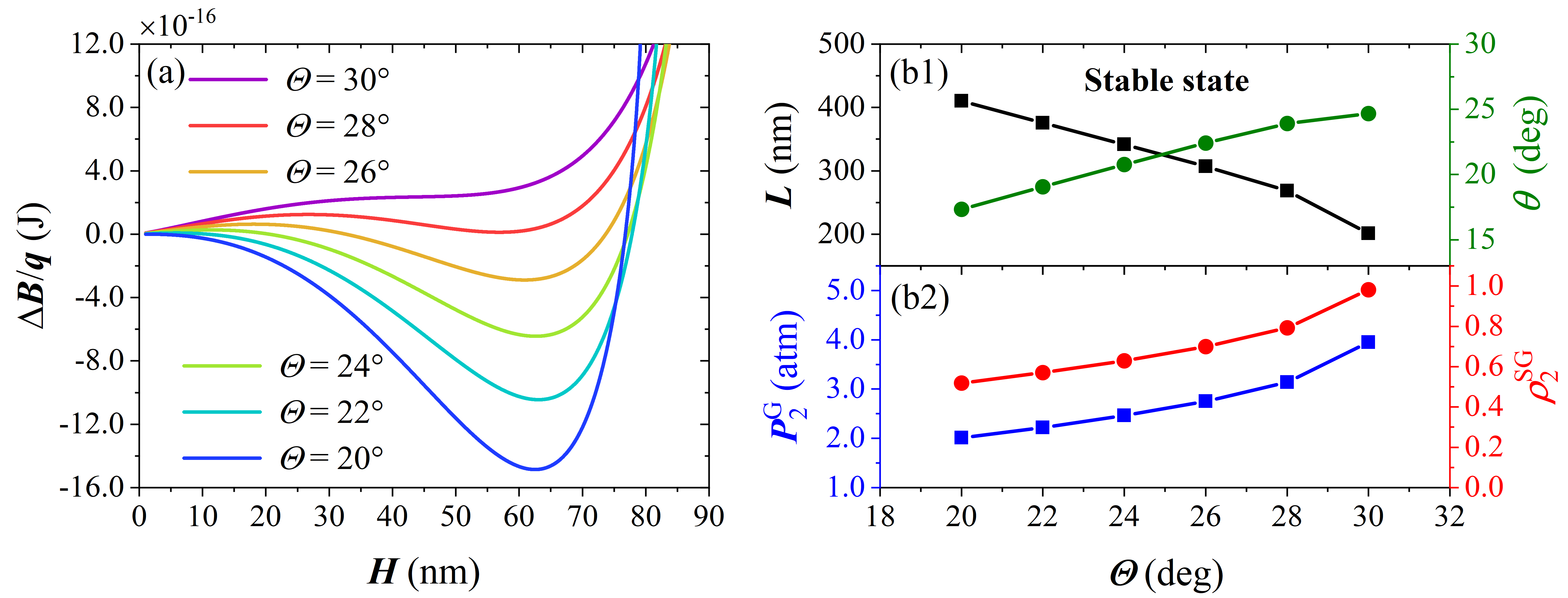}
    \captionsetup{justification=justified, format=plain, width=\textwidth}
    \caption{(a) System's free energy versus bubble height $H$ for different macroscopic contact angles $\mathit{\Theta}$ of substrate surface. (b1) Footprint radius $L$ and contact angle $\theta$, (d2) gas partial pressure $P_2^{\mathrm{G}}$ in bubble phase and adsorbed gas molecular density $\rho_2^{\mathrm{SG}}$ at solid-bubble interface of surface nanobubble in the stable state versus $\mathit{\Theta}$.}
    \label{Fig.5}
\end{figure} 

\autoref{Fig.5} quantitatively analysis the impact of substrate hydrophobicity on the stable states of surface nanobubbles, with $\zeta=4$ and $K=2\times 10^{-7}$. \autoref{Fig.5}(a) demonstrates that reducing the macroscopic gas-side contact angle $\mathit{\Theta}$ of the solid surface causes a downward shift in the free energy curve, suggesting the existence of stable bubbles with lower system free energy. In \autoref{Fig.5}(b1), as $\mathit{\Theta}$ decreases, the solid surface energy also decreases, indicating increased hydrophobicity. This leads to stable nanobubbles with a smaller contact angle and larger footprint radius, resulting in a flatter shape. \autoref{Fig.5}(b2) shows that the flatter shape of bubbles reduces the Laplace pressure, while the decreasing gas partial pressure $P_2^{\mathrm{G}}$ within the bubbles reduces the adsorbed gas molecular density $\rho_2^{\mathrm{SG}}$ at the solid-bubble interfaces. The analysis suggests that hydrophobic solid surfaces promote the formation and stability of SNBs, resulting in bubbles with a flatter shape.

In experiments and simulations \citep{Lohse2015}, the formation and stability of surface nanobubbles are closely influenced by environmental factors, such as increased gas concentration, enhanced hydrophobicity of the substrate surface. These factors can compensate for each other to promote the nucleation and stability of nanobubbles on the solid surface. Therefore, the proposed thermodynamic model can effectively analyze the influence of the surrounding environment in which SNBs exist on their dynamics and stable forms, thereby shedding light on their prolonged existence.

\subsection{Morphology and stability of nanobubbles with different gases}

\begin{figure}
    \centering
    \includegraphics[width=12cm]{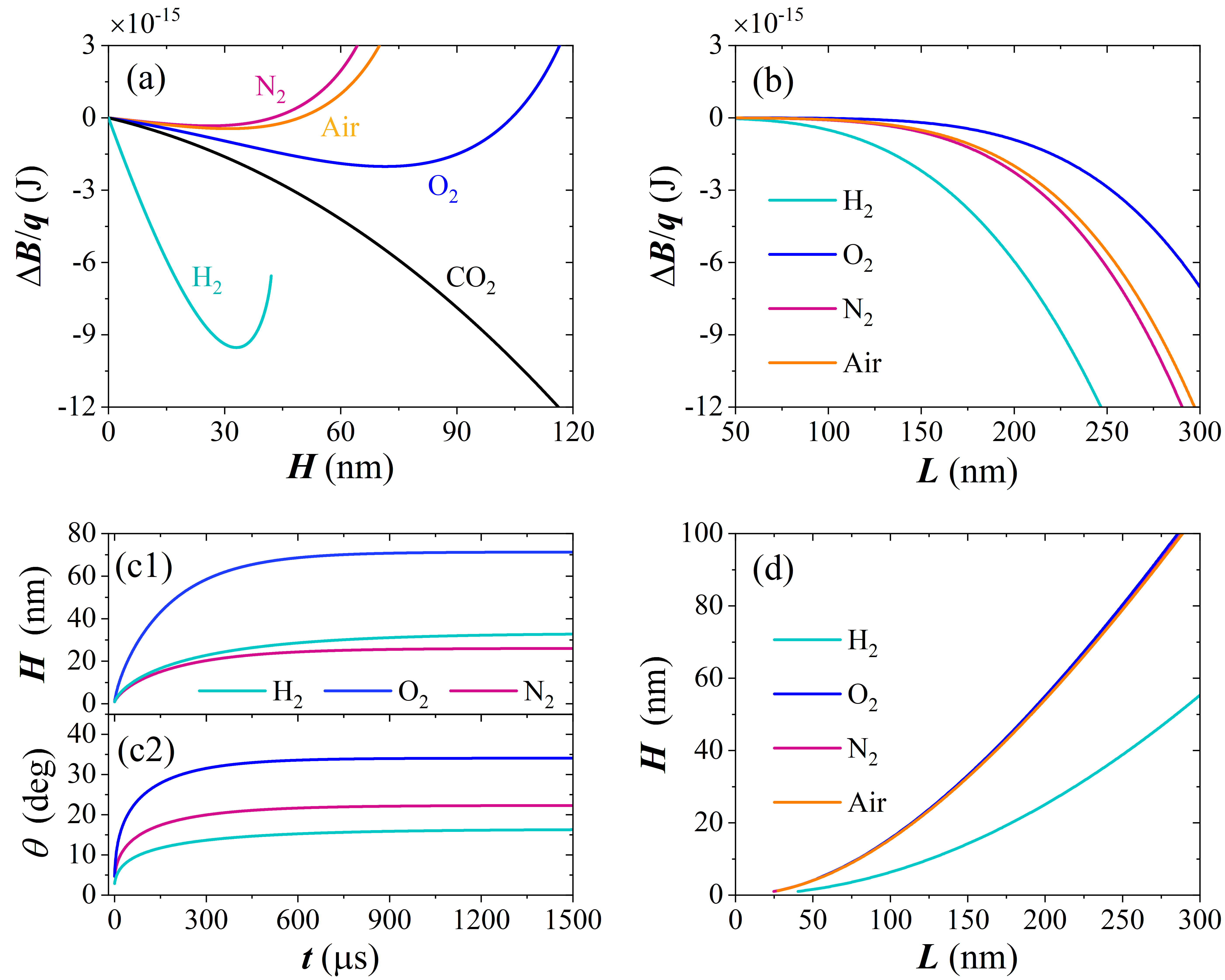}
    \captionsetup{justification=justified, format=plain, width=\textwidth}
    \caption{(a) System's free energy versus bubble heigh $H$ for $\text{H}_2$, $\text{O}_2$, $\text{N}_2$, $\text{CO}_2$, Air nanobubbles on hydrophobic HOPG surfaces. (b) System’s free energy versus footprint radii $L$ of $\text{H}_2$, $\text{O}_2$, $\text{N}_2$, Air nanobubble in stable equilibrium state. (c) Evolution of height $H$ and contact angle $\theta$ of $\text{H}_2$, $\text{O}_2$, $\text{N}_2$ nanobubble over time. (d) Height $H$ versus footprint radii $L$ of $\text{H}_2$, $\text{O}_2$, $\text{N}_2$, Air nanobubble in stable equilibrium state.}
    \label{Fig.6}
\end{figure}

SNB morphology is also closely related to the gas type and properties. In experimental preparations, common gas types include: $\text{H}_2$ (Hydrogen) \citep{Mita2022,German2018}, $\text{O}_2$ (Oxygen) \citep{Mita2022,Yang2009}, $\text{N}_2$ (Nitrogen) \citep{Yang2009,Zhang2006}, $\text{CO}_2$ (Carbon dioxide) \citep{Zhang2008,German2014}, Air \citep{Borkent2009,Zhang2010}. \autoref{Fig.6} compares the characteristics of SNBs on a hydrophobic HOPG (highly oriented pyrolytic graphite, with $\mathit{\Theta}=85^{\circ}$) substrate formed by five different gases in terms of bubble morphology and stability. Due to the differences in gas properties, here, $\text{H}_2$ with $K=1\times 10^{-7}$, and the other with $K=1\times 10^{-5}$. \autoref{Tab1} lists the parameter values of the five gases, including gas-liquid interfacial tension $\gamma^{\mathrm{LG}}$,  the gas solubility $c_{\mathrm{s}}$ in liquid, and the cross-sectional area $b$ of a single gas molecule.

\begin{table}
  \begin{center}
\def~{\hphantom{0}}
  \begin{tabular}{lccccc}
      $\text{Parameter}/\text{Gas}$ & $~~~~\text{H}_2~~~~$&    $~~~~\text{O}_2~~~~$  &  $~~~~\text{N}_2~~~~$&   $~~~~\text{CO}_2~~~~$ &  $~~~~\text{Air}~~~~$ \\[3pt]
       
      $\gamma^{\mathrm{LG}}~(\mathrm{N\cdot m^{-1}})$ & 0.025 & 0.063 & 0.072 & 0.051 & 0.072\\
      $c_{\mathrm{s}}~(\mathrm{kg/nm^3})$ & 0.00016 & 0.048 & 0.017 & 0.819 & 0.022\\
      $b~(\mathrm{nm^2})$  & $1.08\times 10^{-3}$ & $8.04\times 10^{-2}$ &  $7.55\times 10^{-2}$ & $8.52\times 10^{-2}$ & $7.55\times 10^{-2}$\\
  \end{tabular}
  \caption{The gas-liquid interface tension $\gamma^{\mathrm{LG}}$,  the gas solubility $c_{\mathrm{s}}$ in liquid, and the cross-sectional area $b$ of a single adsorbing molecule for $\text{H}_2$, $\text{O}_2$, $\text{N}_2$, $\text{CO}_2$, Air.}
  \label{Tab1}
  \end{center}
\end{table}

In \autoref{Fig.6}(a), the system free energy curves for five gas nanobubbles show that when the system is in equilibrium at the minimum free energy state, the stable $\text{H}_2$ nanobubble is the most stable, with in same initial gas oversaturation $\zeta=6$. The $\text{CO}_2$ nanobubbles are the most unstable, in other words, it is difficult for them to be stable at the nanoscale, which is consistent with the observation that $\text{CO}_2$ SNBs are typically at the micrometer scale in experimental measurements \citep{German2014}. The stable size of $\text{N}_2$, Air, and $\text{O}_2$ nanobubbles increases in sequence, and their stability also increases in sequence. \autoref{Fig.6}(c1) and 6(c2) show the evolution of height and contact angle of $\text{H}_2$, $\text{O}_2$, $\text{N}_2$ nanobubbles from nucleation, growth, to stabilization processes, it can be observed that $\text{O}_2$ nanobubbles take longer to reach larger stable sizes. In addition, under the same footprint radius $L$, stable $\text{H}_2$ nanobubbles present better stability and a flatter morphology, as shown in \autoref{Fig.6}(b) and \autoref{Fig.6}(d). The experiments of \citet{Mita2022} clearly illustrated the characteristic of $\text{H}_2$ SNBs: those with a flatter shape are less mobile and tend to maintain a fixed position. 

\begin{figure}
    \centering
    \includegraphics[width=12cm]{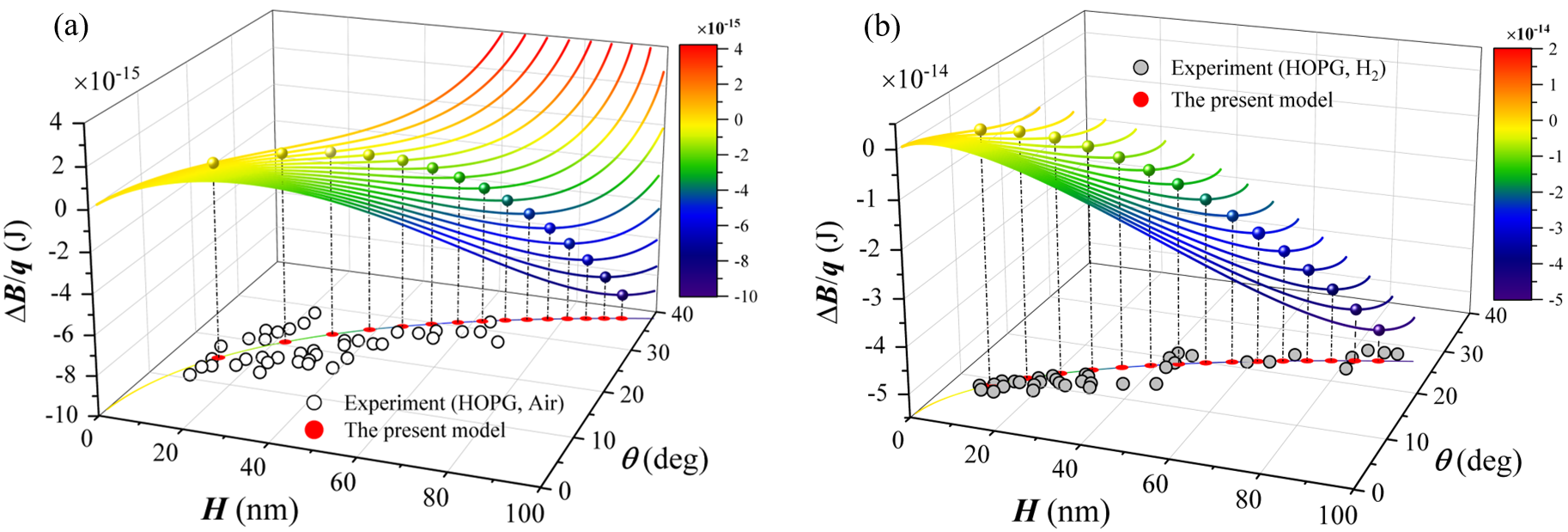}
    \captionsetup{justification=justified, format=plain, width=\textwidth}
    \caption{System's free energy versus bubble height $H$ and contact angle $\theta$ as gas oversaturation increases for (a) Air and (b) $\text{H}_2$ nanobubbles on hydrophobic HOPG surface. The red circles projection of the spheres representing stable bubble states on the $H$-$\theta$ plane. The experimental data of (a) Air (white circles) and (b) $\text{H}_2$ (gray circles) nanobubbles on hydrophobic HOPG surfaces by \citet{Zhang2010}}
    \label{Fig.7}
\end{figure}

\autoref{Fig.7}(a) and \autoref{Fig.7}(b) show the system free energy curves of Air and $\text{H}_2$ (Hydrogen) nanobubbles on the HOPG surface, the spheres mark the stable state of nanobubbles at the minimum system free energy. The changes in the free energy curves indicate that as the gas oversaturation $\zeta$ in the liquid increases, the stable SNBs have larger sizes and stronger stability. It can be observed that the stable state curve of SNBs is connected by red circles, which are projected onto the height $H$ and contact angle $\theta$ plane by the free energy minimum points. In Experiment, \citet{Zhang2010} generated Air and $\text{H}_2$ nanobubbles on the hydrophobic HOPG surface using the alcohol-water exchange method and the electrolysis method, respectively. The size of SNBs distributed on the surface was measured using an atomic force microscope. The experimental data of Air and $\text{H}_2$ nanobubbles are represented by white and gray circles in the $H-\theta$ plane in \autoref{Fig.7}(a) and \autoref{Fig.7}(b), respectively. Comparing the stable state curves and the experimental data for SNBs, the stable morphology of SNBs obtained by our thermodynamic model calculation closely aligns with the experimental phenomenon within an acceptable margin of error. This result strongly supports our theoretical research.

\section{Conclusions}

This study integrates a gas molecular diffusion model with thermodynamic analysis to explore the critical nucleation, evolution and stable state of surface nanobubbles in a closed environment. The gas adsorption effect at the bubble-substrate interface is also considered. The present model ensures that the stable surface nanobubbles unify the mechanical equilibrium at the contact line, the gas diffusion equilibrium at the bubble-liquid interface, and the thermodynamic equilibrium of the whole fluid system, leading to the formation of stable surface nanobubbles.

The analyses suggest that thermodynamic non-equilibrium drives the gas diffusion and the contact line motion of surface nanobubbles. Overcoming the nucleation energy barrier is crucial for bubble formation and growth. The evolved nanobubbles adhere to the contact line mechanical equilibrium and gas diffusion relations. They evolve in a manner that reduces the system's free energy, and are stabilized at the minimum free energy state. Surface nanobubbles are more likely to form and remain stable in environments with hydrophobic substrates and higher gas concentrations in the liquid, which can help explain their prolonged presence. In addition, the stable state curve of surface nanobubbles with minimum free energy predicted by the presented thermodynamic model is consistent with the experimental phenomenon. Under identical same environmental conditions, hydrogen nanobubbles exhibit a flatter shape and better stability.

The present model extends thermodynamic calculations to analyze the evolution of surface nanobubbles, providing valuable insights into the system's dynamic progression. This perspective provides stronger support for the stability explanation of surface nanobubbles and will be particularly valuable for the investigation of fluid motion and interfacial processes. In our upcoming research, based on the established thermodynamic model, we aim to further investigate the stability of surface nanobubbles by considering the gas supersaturation layer \citep{Tan2018} and the resulting variations in the gas-water interfacial tension \citep{Attard2015,Attard2016}.


\backsection[Funding]{This work was supported by the National Natural Science Foundation of China (Nos. 12272100, 11605151), and the Innovation Project of Guangxi Graduate Education (No. YCBZ2024087).}

\backsection[Declaration of interests]{The authors report no conflict of interest.}

\backsection[Author ORCIDs]{Liang Zhao, https://orcid.org/0000-0003-2687-1026;
Binghai Wen, https://orcid.org/0000-0002-0127-3268.}

\bibliographystyle{jfm}
\bibliography{jfm}

\end{document}